\documentclass{elsart}

\usepackage{amsmath,amssymb}
\usepackage{graphicx}

\newcommand{\ee}{$e^+e^-$}               

\newcommand{\pp}{\textit{pp}}
\newcommand{\ppbar}{\textit{p\=p}}
\newcommand{\pt}{p_T}
\def\avg#1{\langle #1 \rangle}  
\def\nbar{\bar n}            
\def\Nbar{\bar N}            
\def\Rs{\sqrt{s}}

\begin{document}

\begin{frontmatter}



\title{Density saturation and the decrease of the normalised
	width of the multiplicity
	distribution in high energy \pp\ collisions}


\author[ist]{Jorge Dias de Deus},
\author[san]{E.G. Ferreiro},
\author[san]{C. Pajares},
\author[tor]{R. Ugoccioni}

\address[ist]{CENTRA and Departamento de F{\'\i}sica (I.S.T.),
  Av. Rovisco Pais, 1049-001 Lisboa, Portugal}
\address[san]{Departamento de F\'{\i}sica de Part\'{\i}culas,
  Universidade de Santiago de Compostela, 15706 Santiago de
	Compostela, Spain}
\address[tor]{Dipartimento di Fisica Teorica and INFN - Sezione di Torino
  Via P. Giuria 1, 10125 Torino, Italy}

\begin{abstract}
It is experimentally observed that the width of the KNO multiplicity
distribution ---or the negative binomial parameter, $1/k$---
for \pp\ collisions, in the energy region $10 \lesssim
\Rs \lesssim 1800$~GeV, is an increasing function of the energy.
We argue that in models with parton or string saturation such trend
will necessary change: at some energy the distribution will start to
become narrower.
In the framework of percolating strings, we have estimated the change
to occur at an energy of the order of 5--10 TeV.
\end{abstract}


\end{frontmatter}

\section{The problem}

Experimentally, in \pp\ and \ppbar\ collisions, not only the average
particle multiplicity is increasing with energy but, as well, the
normalised width of the multiplicity distribution is an increasing
function of energy \cite{NA22:b3+ISR:1}.
In other words, both $\avg{n}$ and $D^2/\avg{n}^2 \equiv (\avg{n^2} - 
\avg{n}^2)/\avg{n}^2$ increase with energy.

The increase of $\avg{n}$ is expected: more particles are produced
when more energy is available.
The growth of $D^2/\avg{n}^2$ is, however, not so obvious.
The general argument seems to be that the increase of energy implies
phase-space opening.
There is more room for partons, strings, clusters or clans, more room
for multiple collisions and different impact parameters contributions,
more room for harder collisions and additional QCD branching
\cite{DPM:1+LundModel+combo:prd+Walker:p+AGLVH:1+pQCD}.
In all these approaches, if there are no limitations, fluctuations
increasingly dominate over average multiplicity: $D^2/\avg{n}^2$
increases and, asymptotically, it may reach a constant value (KNO
scaling \cite{KNO}). 
This kind of behaviour, as pointed out in \cite{Wuhan}, has been observed
long ago in cosmic ray physics \cite{Cosmic}, with 
the negative binomial parameter $k$ stabilising at a
value of the order of 2.5--3, for $\Rs \geq 1$~TeV.

  However, if saturation phenomena occur 
\cite{saturation:1+saturation:2+McLerran:1994ni+McLerran:1994ka+Kovchegov:1996ty,Braun:1992ss}, 
i.e., too many
objects occupy a finite region of interaction, then fluctuations are
severely limited and $D^2/\avg{n}^2$ will start to decrease.
On general grounds, this is not unreasonable.
As the number of intermediate objects (or final particles) is bounded
by $\Rs$ (conservation of energy), while the region of interaction is
constrained by the stronger Froissart bound, $\ln^2\Rs$, there exists
a real possibility for saturation.
It is also well known that, at fixed energy, when triggering on rare
events \cite{components+rare+components:AA} 
or selecting central impact parameter, $b=0$, collisions
\cite{Pajares:1998uq}, the KNO distribution, due to the higher density, is
narrower.

\section{Theoretical framework}

We shall work in the framework of the dual string model
\cite{SFM:1+SFM:2+Braun:Feta+Braun:2000hd+Armesto:SFM}.
Saturation, in this case, results from clustering and percolation of
strings.
In a collision, $N_s$ strings are produced and they may form clusters
with different number of strings (or different sizes).
These clusters act as random, particle emitting, sources.
Following \cite{Pajares:II+Pajares:2004ib}, we write for the multiplicity
distribution, $P(n)$,
\begin{equation}
	P(n) = \int W(X) P(n; X\bar n_1) dX ,  \label{eq:1}
\end{equation}
where $W$ is a normalised weight function, $X$ is a clustering
variable, $P(n; X\bar n_1)$ a convolution function for a cluster of
strings, $\bar n_1$ being the single string average multiplicity.
We assume for $P(n; X\bar n_1)$ a Poisson distribution, this being
consistent with the single string distribution of low energy \ee\
annihilations. 
Note that, as emphasised in \cite{Pajares:II+Pajares:2004ib}, 
an equation similar to 
(\ref{eq:1}), with the same function $W(X)$, can be written for the
inclusive $p_T$ distribution, $f(p_T)$.

From (\ref{eq:1}) we obtain
\begin{equation}
	\avg{n} = \avg{X} \nbar_1   \label{eq:2}
\end{equation}
and
\begin{equation}
	\frac{ 1 }{ k } \equiv \frac{ \avg{n^2}-\avg{n}^2 }{ \avg{n}^2 }
	  - \frac{ 1 }{ \avg{n}}
   = \frac{ \avg{X^2}-\avg{X}^2 }{ \avg{X}^2 } . \label{eq:5}
\end{equation}

Defining now $N_c$ as the number of clusters, $N$ the
number of strings in a cluster and $\Nbar_s$ the average number of
strings, which must satisfy the sum rule
$\avg{N_c} \avg{N} = \bar N_s$,
we shall write
\begin{equation}
	X \equiv \avg{N_c} N ,  \label{eq:7}
\end{equation}
such that, finally, we obtain
\begin{equation}
	\avg{n} = \avg{N_c} \avg{N} \nbar_1 = \Nbar_s \nbar_1 , \label{eq:8}
\end{equation}
and
\begin{equation}
	\frac{ 1 }{ k } = \frac{ \avg{N^2}-\avg{N}^2 }{ \avg{N}^2 } .\label{eq:9}
\end{equation}

The relevant parameter in string clustering (and percolation) is the
transverse density $\eta$,
\begin{equation}
	\eta = \left( \frac{ r_0 }{ R } \right)^2 N_s , \label{eq:10}
\end{equation}
where $r_0$ is the transverse radius of the string (we shall use $r_0=0.2$
fm) and $R$ is the effective radius of the interaction region, with
$R\approx 1$ fm in \pp\ collisions.

It is the parameter $\eta$ that controls clustering: when $N$ strings
cluster, the effective colour charge is not proportional to $N$ but,
due to the vectorial nature of colour summation \cite{Biro:randomsum},
is proportional to $F(\eta)N$, where \cite{Braun:2001us}
\begin{equation}
	F(\eta) = \sqrt{ \frac{ 1 - e^{-\eta} }{\eta } } , \label{eq:11}
\end{equation}
such that, instead of  (\ref{eq:8}), one should write
\begin{equation}
	\avg{n} = F(\eta) \Nbar_s \nbar_1  .  \label{eq:13}
\end{equation}
Equation (\ref{eq:9}) for $1/k$ remains, of course, unchanged.
Note that $F(\eta)$ is a decreasing function of $\eta$.

\section{The toy model}

  In order to show, clearly, the problem of the competition between
the opening of new possibilities and limitations due to saturation,
let us introduce a very simple, and known to everybody, toy model.
Imagine that one has $N_s$ identical coins to distribute over $M$
identical boxes. 
The reader may imagine the distribution of $N_s$
partons or strings over the interaction region of size $M$, or $R^2$
(in units of $r_0^2$).
This is a purely combinatorial problem, without
percolation.

What is the distribution of coins in the boxes? In Ref.\ \cite{F:eta} 
it was shown 
that in the thermodynamical limit only the transverse density
\begin{equation}
	\eta \equiv \frac{N_s}{M} ,
\end{equation}
is relevant, and that the first moments of the distribution 
$P(N)$ of the number 
of strings per cluster (the probability of having a $N$-cluster) 
\begin{equation}
	\avg{N} = \frac{ \eta }{ 1 - \exp(-\eta) } ,   \label{eq:toy8}
\end{equation}
and
\begin{equation}
  K =	\frac{ \avg{N}^2 }{ \avg{N^2} - \avg{N}^2 } 
  = \frac{ \eta }{ 1 - (1+\eta)\exp(-\eta)
                                              } .  \label{eq:toy9}
\end{equation}

If we use this toy model as input for $W(X)$ in (\ref{eq:1}), and
take, as mentioned before, the Poisson distribution for $P(n; \nbar_1)$,
then, see (\ref{eq:9}), one obtains
\begin{equation}
	K = k .
\end{equation}

\begin{figure}
  \begin{center}
  \mbox{\includegraphics[width=\textwidth]{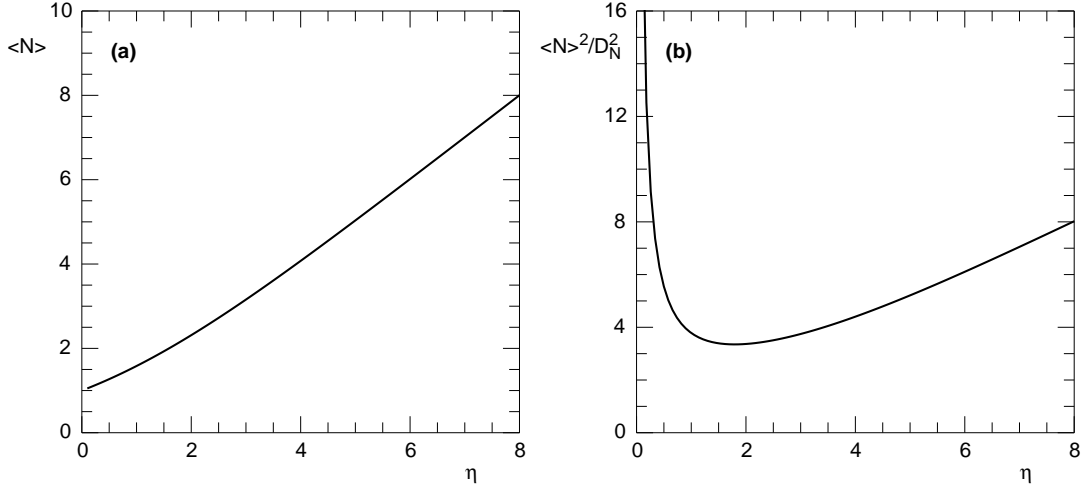}}
  \end{center}
  \caption{Plots of \textbf{(a)} the average number of strings per
		cluster, Eq.~(\ref{eq:toy8}) and 
		\textbf{(b)} of the inverse of its normalised variance, 
		or $k$, Eq.~(\ref{eq:toy9}),
    in the toy model, vs.\ the string density $\eta$.}\label{fig:1}
  \end{figure}

In Figure~\ref{fig:1} we present plots of (\ref{eq:toy8}) and (\ref{eq:toy9}).
The average multiplicity is, naturally, a growing function of the density.
The parameter $k$ goes to infinity at small $\eta$ (one just has 1-clusters),
decreases and reaches a minimum, $k \approx 3.3$, at $\eta \approx
1.8$, then moves to infinity again as $\eta\to\infty$ (the
distribution is dominated by $N_s/M$-clusters).
We have checked by Monte Carlo simulation that the distribution $P(N)$
of the model is similar to a gamma distribution.

\section{The physics}

In order to make contact with real physics, we need to estimate the
dependence of the parameter $k$, experimentally measured in \pp\
collisions as a function of the energy, on the density parameter
$\eta$.
In other words, we have to find the energy dependence of $\eta$.
As $\eta$, see (\ref{eq:10}), depends on $N_s$ and $R^2$, we need to
find the energy dependence of $N_s$ and $R^2$.

For the \pp\ average charged particle multiplicity in the dual string
model, we shall write \cite{RU:dNdeta+RU:dNdeta:2}:
\begin{equation}
	\avg{ n(\Rs) } = F(\eta) \nbar_1 \left[
	     2 + (N_s(\Rs) - 2)\alpha \right] ,       \label{eq:phys12}
\end{equation}
where $\nbar_1$ is the single valence string average multiplicity
(assumed constant), $\alpha$ the sea string multiplicity reduction
factor ($\alpha < 1$).
Notice that in (\ref{eq:phys12}) one always has 2 valence strings and $N_s-2$
sea strings.
At low energy, $\Rs \approx 10$~GeV, experimentally $\avg{n}
\approx 4$, and theoretically $N_s \approx 2$ (the two valence
strings) and of course $F(\eta)\approx 1$ and $\nbar_1 \approx 2$.
Notice that (\ref{eq:phys12}) is a natural generalisation of
(\ref{eq:13}) when there are two kinds of strings.

As the slope parameter $B(\Rs)$  is a direct measure of the
square of the interaction radius $R^2(\Rs)$, we write, from (\ref{eq:10}),
\begin{equation}
	\frac{ N_s(\Rs) }{ \eta } = \left( \frac{ R(\Rs) }{ r_0 } \right)^2
	    = c B(\Rs) ,                           \label{eq:phys13}
\end{equation}
with $c \approx 1.6 ~{\text{GeV}}^2 {\text{fm}}^2$, such that at $\Rs
\approx 10$~GeV, $R\approx 1$~fm and $B\approx 10~{\text{GeV}}^{-2}$.

By using equations (\ref{eq:phys12}) and (\ref{eq:phys13}), 
and experimental information
on $\avg{n(\Rs)}$ \cite{ndata} and on $B(\Rs)$ \cite{Bdata}, we can in
principle construct $N_s(\Rs)$ and $\eta(\Rs)$:
from $\eta(\Rs)$ we obtain $k(\Rs)$, Eq.~(\ref{eq:toy9}).

\begin{figure}
  \begin{center}
  \mbox{\includegraphics[width=0.9\textwidth]{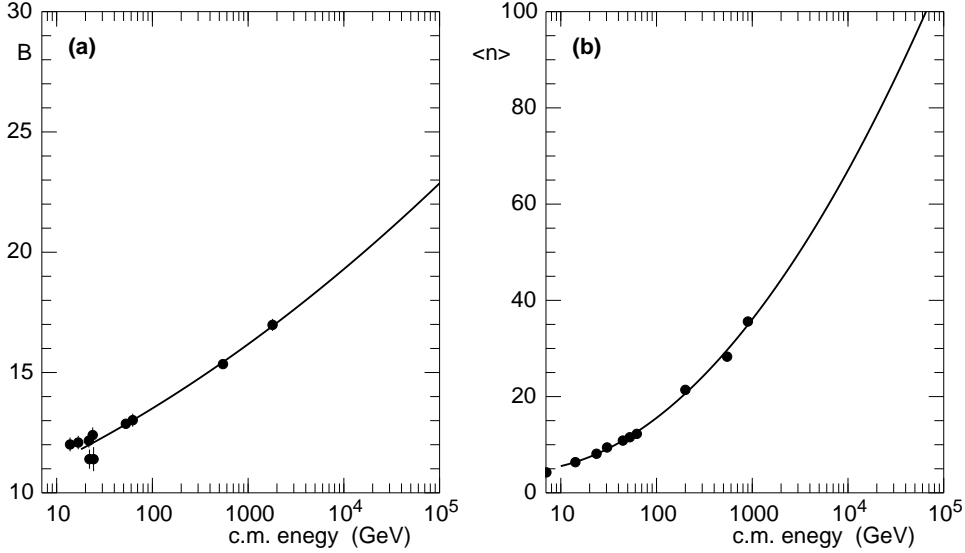}}
  \end{center}
  \caption{\textbf{(a)} Slope parameter $B$ (GeV$^{-2}$) vs.\ c.m.\
		energy (data points from 
		\cite{Bdata}), parametrised by $B(\Rs) = 9.52 + 0.671 \ln\Rs + 
		0.0424\ln^2 \Rs$. \textbf{(b)}
    Average multiplicity (data points from \cite{ndata}) 
    vs.\ c.m.\ energy, parametrised by $\avg{n(\Rs)} = 5.99 - 2.46 \ln\Rs 
    + 0.987 \ln^2\Rs$.}\label{fig:2}
  \end{figure}

In Fig.~\ref{fig:2}a we show our fit to $B(\Rs)$ and in
Fig.~\ref{fig:2}b our fit to $\avg{n(\Rs)}$.
These parameterisations are then used in Eq.~(\ref{eq:phys13}) and
Eq.~(\ref{eq:phys12}). 

\begin{figure}
  \begin{center}
  \mbox{\includegraphics[height=0.65\textheight]{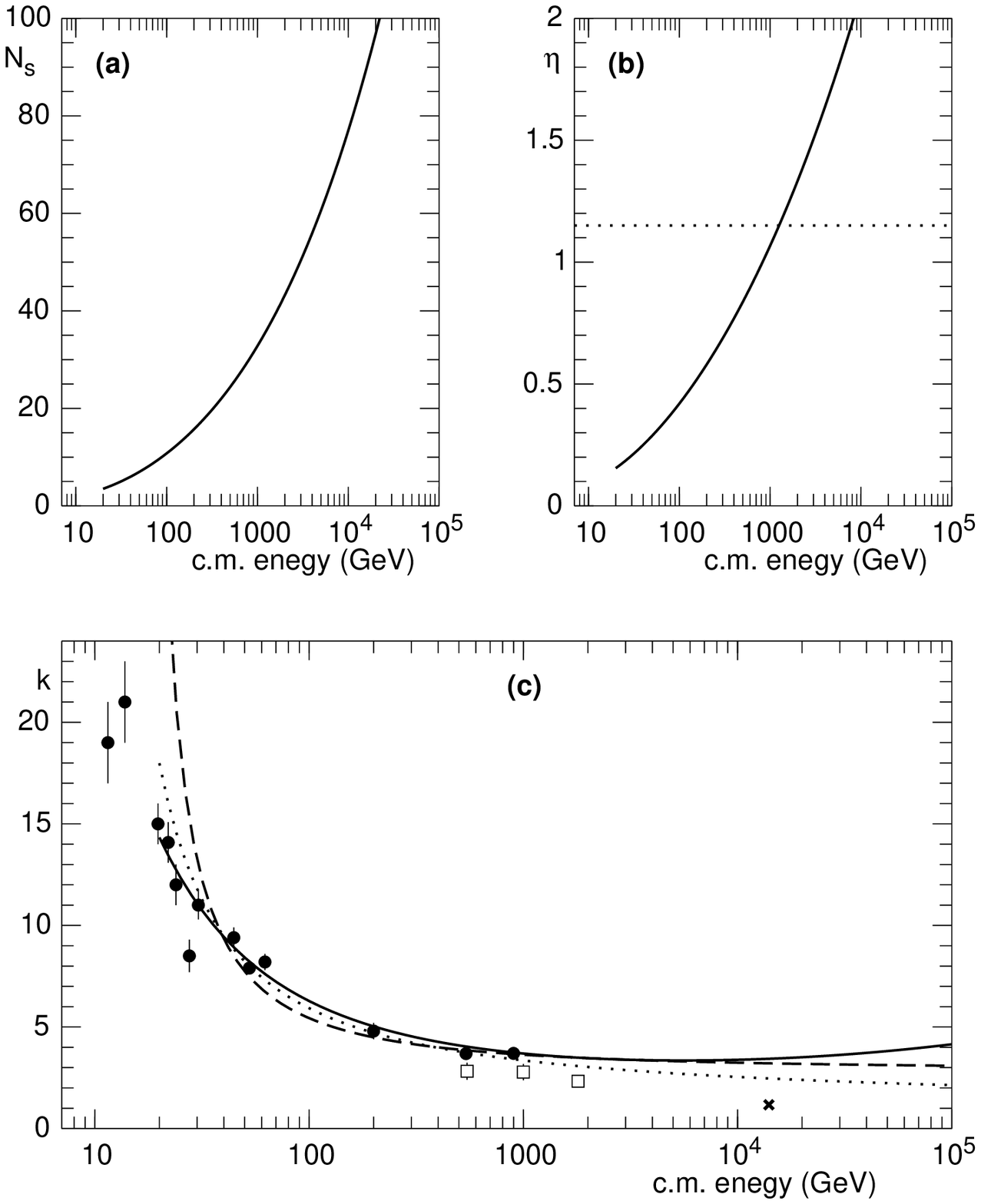}}
  \end{center}
  \caption{\textbf{(a)} $N_s$ vs c.m.\ energy;
		\textbf{(b)} $\eta$ vs c.m.\ energy; 
	  \textbf{(c)} $k$ vs c.m.\ energy; experimental points from ISR and
  UA5 (filled points, \cite{ndata}) 
	and E735 (open squares, \cite{Walker}).
  The values of the parameters used [see (\ref{eq:phys12}) and
		(\ref{eq:phys13})] are: $\alpha=0.45$, $\nbar_1=2.9$ and 
	$c=1.9~{\text{GeV}}^2 {\text{fm}}^2$.
  The dashed line is a fit using Wr\'oblewski regularity,
	$D = 0.59\avg{n} - 1.4$; the dotted line is a fit using
	$k^{-1} = 0.154\ln\avg{n} - 0.25$
  (both fits have been carried out
  considering only data with $\sqrt{s} \ge 30$ GeV.)
  The cross is the prediction by the Pythia Monte Carlo 
  with the most recent
  parameters optimisation \cite{Moraes}.}\label{fig:3}
  \end{figure}

Intuitively, one expects the number of strings, $N_s$, to increase
with $\Rs$.
The results from our fits are shown in Fig.~\ref{fig:3}: $N_s$
increases with energy and, within a factor 2, is not substantially
different from Monte Carlo direct calculations \cite{Pajares:perc}.
Naively, as $\eta \sim N_s$, one also expects $\eta$ to grow with
energy.
However, as the interaction area also grows with energy, the situation
is not so clear.
From (\ref{eq:phys12}) and (\ref{eq:phys13}) it is easily seen that the $\eta$
dependence on $\Rs$, $\eta(\Rs)$, is determined by the ratio
$\avg{n(\Rs)} / B(\Rs)$.
In Figure \ref{fig:3}b we present $\eta(\Rs)$;
in the figure is shown as well the percolation threshold: $\eta_c
\simeq 1.15$.

Finally, in Fig.~\ref{fig:3}c, we show the dependence of $k$ on energy,
see (\ref{eq:toy9}) and Fig.~\ref{fig:3}b, in comparison with experimental
points.
If the ``coins in boxes'' model reflects the physical situation, we
expect $k$ to reach a minimum at $\Rs \approx 6$ TeV, in the LHC
energy region.
The behaviour we predict is qualitatively different from what is
expected on the basis of extrapolations of other phenomenological
models:
for instance, consider the Wr\'oblewski regularity \cite{Wrobl}: $D = a
\avg{n} + b$; this leads to an asymptotic ($\avg{n}\to\infty$) 
KNO scaling, thus to a levelling off of $k$ (see dashed line in
Fig.~\ref{fig:3}c.)
Other proposals (e.g.\ $k^{-1} = a + b\ln\avg{n}$ in
\cite{Singh:1989fd}, and the Pythia Monte Carlo \cite{Moraes}, 
respectively dotted line and cross in Fig.~\ref{fig:3}c) 
give a monotonically decreasing $k$;
for a discussion of the situation in which $k$ continues to decrease
at high energy and reaches values smaller than 1, see \cite{RU:NewPhysics}.

\begin{figure}
  \begin{center}
  \mbox{\includegraphics[width=0.9\textwidth,height=0.8\textheight]{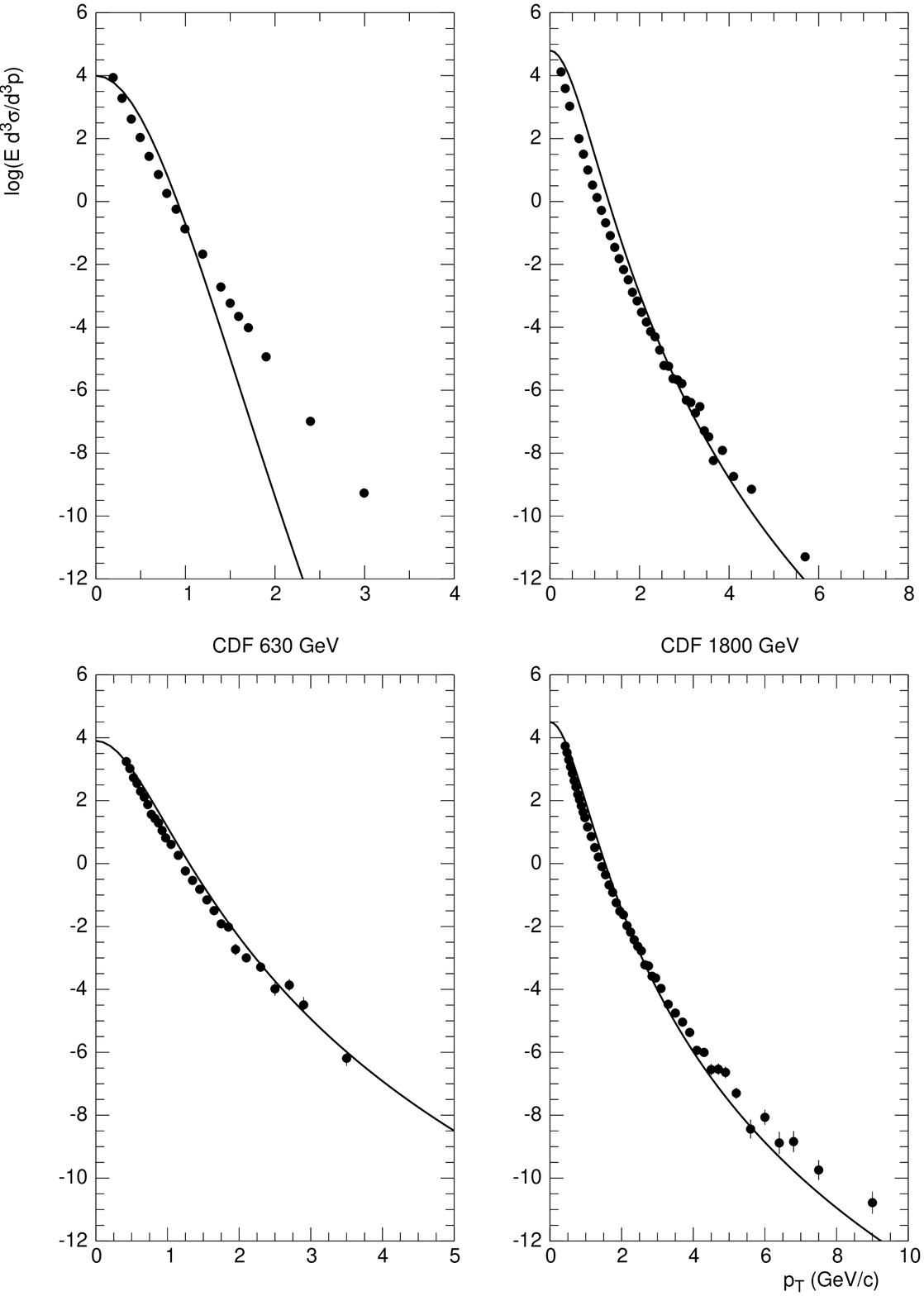}}
  \end{center}
  \caption{$\pt$ distributions (not normalised) at several c.m.\ energies, (from
     \cite{ndata,Albajar:1990an+Abe:1988yu}) fitted with the power-law 
	   $A (1 + b \pt^2)^{-k}$, using, for each energy,
		 the respective value of $k$ obtained from the multiplicity distribution
     (from Fig.~\ref{fig:3}), $A$ and $b$ being free 
     parameters.}\label{fig:4}
  \end{figure}

\section{Conclusions}

We have argued that the energy behaviour of the parameter $k$ (or
$\avg{N}^2/D_N^2$) of the multiplicity distribution in \pp\ collisions
is sensitive to the energy dependence of the parton/string transverse
density $\eta$.
If impact parameter overcrowding never occurs ($\eta$ is always small)
$k$ will decrease with energy, reaching, eventually, some asymptotic constant
value.
If overcrowding takes place ($\eta$ increases rapidly with energy) then $k$
necessarily will change its behaviour and start increasing with
energy.
The turning point may occur at an energy of the order of 5--10~TeV.

It was shown in \cite{Pajares:I,Pajares:II+Pajares:2004ib} 
that the parameter $k$ of the multiplicity
distribution is the parameter $k$ of the $\pt$-distribution:
$(1 + b \pt^2)^{-k}$.
This parameter $k$ is also a decreasing function of energy in the $\Rs
\leq 1.8$~TeV region.
In Figure~\ref{fig:4} we have compared 
experimental data on $\pt$ distributions to the mentioned
parameterisation, with $k$ given by the multiplicity distribution values at the 
same energy. The agreement is reasonable, and improving with energy.

\section{Acknowledgements}

R. Ugoccioni would like to thank  for discussions A. Prokudin on
elastic collisions and M. Monteno on Monte Carlo tuning.
E.G. Ferreiro and C. Pajares thank the financial support of the CICYT of
Spain through the contract FPA2002-01161 and
of Xunta de Galicia through the contract PGIDIT03PXIC20612PN


\begin{thebibliography}{10}

\bibitem{NA22:b3+ISR:1}
{M.\ Adamus et al., NA22 Collaboration}, Z. Phys.\ {\bf C32},  475  (1986);
{A.~Breakstone et al.}, Il Nuovo Cimento {\bf 102A},  1199  (1989).

\bibitem{DPM:1+LundModel+combo:prd+Walker:p+AGLVH:1+pQCD}
{A. Capella, U.P. Sukhatme, C.I. Tan and J. Tr\^an Thanh V\^an}, Physics
  Reports {\bf 236},  225  (1994);
{B. Andersson}, {\em The Lund Model} (Cambridge University Press, Cambridge,
  U.K., 1996);
{A.~Giovannini and R.~Ugoccioni}, Phys.\ Rev.\ D {\bf 59},  094020  (1999);
{S.G.~Matinyan and W.D.~Walker}, Phys.\ Rev.\ {\bf D59},  034022  (1998);
{A. Giovannini and L. Van Hove}, Z. Phys.\ C {\bf 30},  391  (1986);
{Yu.L. Dokshitzer, V.A. Khoze, A.H. Mueller and S.I. Troyan}, {\em Basics of
  perturbative QCD} (Editions Fronti\`eres, Gif-sur-Yvette, 1991).

\bibitem{KNO}
{Z.\ Koba, H.B.\ Nielsen and P.\ Olesen}, Nucl.\ Phys.\ {\bf B40},  317
  (1972).

\bibitem{Wuhan}
{A.\ Giovannini, invited talk},  in {\em XXI International Symposium on
  Multiparticle Dynamics}, edited by {Liu Lianshou and Wu Yuanfang} (World
  Scientific, Singapore, 1992), p.\ 285.

\bibitem{Cosmic}
{P.K.\ MacKeown and A.W.\ Wolfendale}, Proc.\ Phys.\ Soc.\ {\bf 89},  553
  (1966).

\bibitem{saturation:1+saturation:2+McLerran:1994ni+McLerran:1994ka+Kovchegov:1996ty}
{L.V. Gribov, E.M. Levin and M.G. Ryskin}, Physics Reports {\bf 100},  1
  (1983);
{A.H. Mueller and J. Qiu}, Nucl.\ Phys.\ {\bf B268},  427  (1986);
{L. McLerran and R. Venugopalan}, Phys.\ Rev.\ D {\bf 49},  2233  (1994);
Phys.\ Rev.\ D {\bf 49},  3352  (1994);
{Yu.V. Kovchegov}, Phys.\ Rev.\ D {\bf 54},  5463  (1996).

\bibitem{Braun:1992ss}
{M.A. Braun and C. Pajares}, Phys.\ Lett.\ B {\bf 287},  154  (1992).

\bibitem{components+rare+components:AA}
{J. Dias de Deus, C. Pajares and C.A. Salgado}, Phys.\ Lett.\ {\bf B407},  335
  (1997);
Phys.\ Lett.\ {\bf B408},  417  (1997);
Phys.\ Lett.\ {\bf B409},  474  (1997).

\bibitem{Pajares:1998uq}
{C. Pajares and Yu.M. Shabelski}, Phys. Atom. Nucl. {\bf 63},  908  (2000).

\bibitem{SFM:1+SFM:2+Braun:Feta+Braun:2000hd+Armesto:SFM}
{N.S. Amelin, M.A. Braun and C. Pajares}, Phys.\ Lett.\ {\bf B306},  312
  (1993);
Z. Phys.\ {\bf C63},  507  (1994);
{M.A. Braun and C. Pajares}, Eur.\ Phys.\ J.\ {\bf C16},  349  (2000);
Phys.\ Rev.\ Lett.\ {\bf 85},  4864  (2000);
{N. Armesto and C. Pajares}, {Int.\ J.\ Mod.\ Phys.\ } {\bf A15},  2019
  (2000).

\bibitem{Pajares:II+Pajares:2004ib}
{J. Dias de Deus, E.G. Ferreiro, C. Pajares and R. Ugoccioni}, preprint
  hep-ph/0304068;
{C. Pajares}, Acta Phys.\ Pol.\ B {\bf 35},  153  (2004).

\bibitem{Biro:randomsum}
{T.S. Biro, H.B. Nielsen and J. Knoll}, Nucl.\ Phys.\ {\bf B245},  449  (1984).

\bibitem{Braun:2001us}
{M.A. Braun, F. Del Moral and C. Pajares}, Phys.\ Rev.\ C {\bf 65},  024907
  (2002).

\bibitem{F:eta}
{M.A. Braun, C. Pajares and J. Ranft}, {Int.\ J.\ Mod.\ Phys.\ } {\bf A14},
  2689  (1999).

\bibitem{RU:dNdeta+RU:dNdeta:2}
{J. Dias de Deus and R. Ugoccioni}, Phys.\ Lett.\ {\bf B491},  253  (2000);
Phys.\ Lett.\ {\bf B494},  53  (2000).

\bibitem{ndata}
{G.~Giacomelli and M.~Jacob}, Physics Reports {\bf 55},  1  (1979);
{G.J.\ Alner et al.\ (UA5 Collaboration)}, Physics Reports {\bf 154},  247
  (1987).

\bibitem{Bdata}
{J.P. Burq et al.\ (NA8 Collaboration)}, Phys.\ Lett.\ B {\bf 109},  124
  (1982);
{M. Adamus et al.\ (NA22 Collaboration)}, Phys.\ Lett.\ B {\bf 186},  223
  (1987);
{N. Amos et al.\ (ISR)}, Nucl.\ Phys.\ B {\bf 262},  689  (1985);
{R.E. Breedon et al.\ (UA6 Collaboration)}, Phys.\ Lett.\ B {\bf 216},  459
  (1989);
{F. Abe et al.\ (CDF Collaboration)}, Phys.\ Rev.\ D {\bf 50},  5518  (1994).

\bibitem{Walker}
{T. Alexopoulos et al. (E735 Coll.)}, Phys.\ Lett.\ {\bf B435},  453  (1998).

\bibitem{Wrobl}
{A.K.\ Wr\'oblewski}, Acta Phys.\ Pol.\ {\bf B4},  857  (1973).

\bibitem{Singh:1989fd}
C.~P. Singh and A.~N. Kamal, Phys.\ Lett.\ B {\bf 237},  284  (1990).

\bibitem{RU:NewPhysics}
{A. Giovannini and R. Ugoccioni}, Phys.\ Rev.\ D {\bf 68},  034009
(2003).

\bibitem{Moraes}
T. Sj{\"o}strand {\it et~al.}, Comput. Phys. Commun. {\bf 135},  238
(2001); parameters tuned as detailed in C.M. Buttar {\it et~al.}, 
Acta Phys.\ Pol.\ {\bf B35}, 433 (2004); A. Moraes  in
\textit{Workshop MC4LHC} (CERN, 31 July 2003), 
and \textit{Physics at LHC} (Prague 2003).
The actual Pythia 6.214 point in the figure is from
M.\ Monteno (private communication).

\bibitem{Pajares:perc}
{N. Armesto, M.A. Braun, E.G. Ferreiro and C. Pajares}, Phys.\ Rev.\ Lett.\
  {\bf 77},  3736  (1996).

\bibitem{Albajar:1990an+Abe:1988yu}
{C. Albajar et al.\ (UA1 Collaboration)}, Nucl.\ Phys.\ B {\bf 335},  261
  (1990);
{F. Abe et al.\ (CDF Coll.)}, Phys.\ Rev.\ Lett.\ {\bf 61},  1819  (1988).

\bibitem{Pajares:I}
{J. Dias de Deus, E.G. Ferreiro, C. Pajares and R. Ugoccioni}, Phys.\ Lett.\ B
  {\bf 581},  156  (2004).

\end{thebibliography}

\end{document}